# Nonlinear Optical Observation of Coherent Acoustic Dirac Plasmons in Thin-Film Topological Insulators

Yuri D. Glinka,[1,2]* Sercan Babakiray,[1] Trent A. Johnson,[1] Mikel B. Holcomb,[1] and David Lederman[1,3]

[1]Department of Physics and Astronomy, West Virginia University, Morgantown, WV 26506-6315, USA, [2]Institute of Physics, National Academy of Sciences of Ukraine, Kiev 03028, Ukraine, [3]Department of Physics, University of California, Santa Cruz, CA 90564, USA

**Abstract**

**Low-energy collective electronic excitations exhibiting sound-like linear dispersion have been intensively studied both experimentally and theoretically for a long time. However, coherent acoustic plasmon modes appearing in time-domain measurements are rarely observed due to Landau damping by the single-particle continua. Here we report on the observation of coherent acoustic Dirac plasmon (CADP) modes excited in indirectly (electrostatically) opposite-surface coupled films of the topological insulator $Bi_2Se_3$. Using transient second harmonic generation, a technique capable of independently monitoring the in-plane and out-of-plane electron dynamics in the films, the GHz-range oscillations were observed without corresponding oscillations in the transient reflectivity. These oscillations were assigned to the transverse magnetic and transverse electric guided CADP modes induced by the evanescent guided Lamb acoustic waves and remained Landau undamped due to fermion tunneling between the opposite-surface Dirac states.**

*The corresponding author: Yuri D. Glinka (yuridglinka@yahoo.com)

**Introduction**

Heavily doped and highly photo-excited three-dimensional (3D) semiconductors with high carrier densities (~$10^{17}$-$10^{19}$ cm$^{-3}$) can sustain quantized longitudinal collective oscillations (plasmons) resulting from Coulomb interactions. These oscillations occur at the plasma frequency $\omega_\text{p} = \sqrt{ne^2/m\varepsilon_0}$ at vanishing wave vectors ($q \to 0$), where $n$, $e$, $m$, and $\varepsilon_0$ are density, charge, mass of electrons and the permittivity of free space, respectively. Consequently, $\omega_\text{p}$ lies in the far-infrared/infrared range since the carrier density is low compared to 3D metals (~$10^{23}$ cm$^{-3}$) where $\omega_\text{p}$ lies in the ultraviolet range. These plasmon modes and their interactions with lattice phonons in semiconductors have been extensively studied using optical spectroscopy[1-8] and were usually referred to as conventional 3D plasmons with a parabolic dispersion $\omega(q) = \omega_\text{p} + 3\upsilon_\text{F}^2 q^2/10\omega_\text{p}$, where $\upsilon_\text{F}$ is the Fermi velocity. In contrast, the conventional 3D surface plasmon (polariton) is a self-sustaining oscillation of the semi-infinite electron gas whose surface plasmon frequency is $\omega_\text{sp} = \omega_\text{p}/\sqrt{1+\varepsilon_\text{sur}}$, where $\varepsilon_\text{sur}$ is the dielectric constant of the surrounding medium.[9] The resulting frequencies of conventional 3D surface plasmons in metals and conventional localized 3D surface plasmons in metal nanostructures may appear in the visible region, allowing for numerous photonic applications.[10]

3D metals can also support monolayer-thick surface electronic states that form a two-dimensional (2D) electron-density layer.[11] 2D plasmons with energy vanishing as $q \to 0$ have been observed in charge inversion layers of *p*-type Si(100),[12] quasi-2D systems of monoatomic metal layers on semiconductors,[13-15] and extrinsic free-standing graphene.[16] Initially it was proposed that 2D electron layer with density $n_\text{2D}$ could support only high-energy optical plasmons (OPs) with $\omega_\text{OP}(q) = \sqrt{n_\text{2D}e^2 q/2m\varepsilon_0}$ dispersion,[12-17] because low-energy acoustic plasmons (APs) with sound-like $\omega_\text{AP}(q) = \upsilon_\text{p} q$ dispersion (where $\upsilon_\text{p}$ is the sound phase velocity) are expected to be Landau damped by the underlying 3D electrons.[18] Alternatively, the excitation of APs has been theoretically predicted for two-component 3D plasmas with a large difference in masses of the light and heavy particles[19,20] and for two-component 2D spatially separated plasmas.[18] In both cases, APs are expected to behave as high-frequency modes with plasmon energy ($\hbar\omega_{AP}$) dispersing away from the single-particle continua. APs are also expected to be excited on 3D metal surfaces due to the nonlocality of the 3D dynamical response.[21] Because APs may mediate the Cooper pairing of particles, the AP concept has been used in theoretical studies of superconductivity in high-Tc cuprates,[22,23] $MgB_2$ structures,[24] and layered structures.[25,26]

The frequency-domain AP features were observed using transmission spectroscopy in charge inversion layers ($\hbar\omega_{AP}$ ~1-3 meV),[12] using Raman spectroscopy in the photoexcited 3D plasma of GaAs ($\hbar\omega_{AP}$ ~5 meV),[27] and using angle-resolved high resolution electron energy loss spectroscopy on bare surfaces of Be(0001) ($\hbar\omega_{AP}$ ~300-2000 meV),[11] Cu(111) ($\hbar\omega_{AP}$ ~180-800 meV),[28] Au(111) ($\hbar\omega_{AP}$ ~275-2200 meV),[29] epitaxial graphene on Pt(111) surface ($\hbar\omega_{AP}$ ~300-3000 meV),[30] and $Bi_2Se_3$ single crystals ($\hbar\omega_{AP}$ ~100-200 meV).[31] Time-domain detection of coherent APs has been reported only for GaAs slabs using a photocurrent autocorrelation technique[32] and transient reflectivity (TR) exploiting two-pump beams[33] ($\hbar\omega_{AP}$ ~0.5-7 meV).

The topological insulator (TI) $Bi_2Se_3$, having an insulating gap in the bulk (~0.3 eV) and intrinsic metallic-type Dirac surface states (SS),[34,35] resembles the artificially created quasi-2D systems of monoatomic metal layers on semiconductors,[13-15] thus offering similar conditions for 2D plasmon excitation. In principle, low-energy (GHz-range) coherent APs for the two component Dirac fermions [i.e. coherent acoustic Dirac plasmons (CADPs)] may exist due to a very anisotropic mass tensor of Bi-related materials (such as Bi and $Bi_2Te_3$).[36] However, this tensor for $Bi_2Se_3$ is almost isotropic,[37] causing the CADPs to be sharply Landau damped by the high-density single-particle continua resulting from natural *n*-doping. This statement is also applicable to thin $Bi_2Se_3$ films thicker than $d = 15$ nm, despite being thinner, two THz-range (1 THz ~4.14 meV) plasmon modes ("optical" and "acoustic") can be excited, similarly to double layer structures, such as bilayer graphene.[38-41] The latter plasmon modes can be detected using the nonlinear optical techniques of four-wave mixing, difference frequency generation, and stimulated Raman scattering.[41-43] These techniques deal with anisotropy between in-plane and out-of-plane tensor components of nonlinear susceptibilities and hence avoids the photon-plasmon momentum mismatch that in the linear optical techniques is usually compensated using the waveguide evanescent Otto/Kretschmann prism couplers or grating structures.[44,45] Because the phase-matched plasmon frequency in the latter case is exclusively determined by the micro-ribbon width and the grating period,[46,47] the linear optical techniques seem to be less flexible and less sensitive to the actual (intrinsic) plasmon dynamics in the films, especially when $d$ matches the range corresponding to direct (wave-functions overlap) and indirect (electrostatic) opposite-surface coupling ($d < 6$ nm and $d < 15$ nm, respectively).[41,48-51] It is worth noting that the term "acoustic" used for the double layer structures is not always related to the excitation of real acoustic waves and is used only to distinguish between the antisymmetric ("acoustic") and symmetric ("optical") modes. These considerations indicate that unique conditions are required to observe CADPs in thin $Bi_2Se_3$ films.

Here we report on the observation of transverse magnetic (TM) and transverse electric (TE) guided CADP modes in thin films of the TI $Bi_2Se_3$. A combined ultrafast linear-nonlinear optical pump-probe technique was used to simultaneously measure TR and transient second harmonic generation (TSHG) in the reflection geometry exploiting various incident/outgoing probe light polarizations.[52,53] In addition, TSHG rotational anisotropy (TSHGRA) was measured as a function of pump-to-probe delay time. We show that TSHG and TSHGRA techniques allow for independent monitoring of temporal electron dynamics confined to the topmost Se atomic layer along the in-plane and out-of-plane directions of the films, thus providing evidence for the excitation of TM and TE guided CADP modes, respectively. We also demonstrate that the excitation of Landau undamped CADPs occurs at a unique condition which has been theoretically predicted for the coupled bilayer structures[54] and can be associated with fermion tunneling between the opposite-surface Dirac SS. For a driving source of TM and TE guided CADP modes, we consider the strong resonant coupling of Dirac fermions to evanescent (non-propagating) guided Lamb waves excited within the same relaxation process.[55] This kind of acousto-plasmonic control is known to occur in complex nanostructures,[56] while the resonance-type acoustic-plasmon-to-acoustic-phonon coupling has been considered theoretically only for Bi.[36]

## Results

**The origin of oscillations in TR and TSHG signals.** The TR and TSHG traces measured simultaneously for the 10 nm thick $Bi_2Se_3$ film at various pump/probe powers and in the $P_{in}$ - $S_{pump}$ - $P_{out}$ and $S_{in}$ - $P_{pump}$ - $S_{out}$ polarization geometries are shown in Figs. 1 and 2, respectively, where P and S denote the polarization of the incident probe and pump laser beams ("in" and "pump") and the outgoing reflected fundamental/SHG probe beam ("out") in the plane of incidence (*xz*) and in the plane of the film (*xy*), respectively. The positive sign of the TSHG response refers to the quadratic form of the SHG intensity that is governed by both the purely surface and depletion-electric-field-induced bulk contributions because of the centrosymmetric nature of $Bi_2Se_3$.[52,53,57,58] The SHG intensity can hence be expressed as

$$I(2\omega) = A \left| \chi_{ijk}^{(S)} + \chi_{ijkl}^{(B)} E_{dc}^{dep} \right|^2 I^2(\omega), \tag{1}$$

where $A$ is a proportionality coefficient, $\chi_{ijk}^{(S)}$ and $\chi_{ijkl}^{(B)}$ are the components of the surface second-order and bulk third-order susceptibility tensors, $E_{dc}^{dep}$ is the dc(direct current)-type depletion electric field of frequency much lower than that of the driving electric field of the incident light $E(\omega)$, and $I(\omega) = E(\omega)E(\omega)$ is the intensity of the incident probe beam. Consequently, the TSHG response monitors the variations of the $E_{dc}^{dep}$ strength due to the dynamical spatial redistribution of photoexcited carriers in close proximity to the surface. Specifically, for the actual free carrier density in the film, $E_{dc}^{dep}$ extends inward toward the film by ~6.5 nm, thus being comparable to the incident and SHG light penetration depth (~10 and ~7 nm, respectively).[41,48-50,58-62] This field can also be modulated by the periodic electric field associated with CADPs. Owing to indirect opposite-surface coupling in the film,[48-50] this modulation will always contribute to the nonlinear susceptibilities independently of whether the TM or TE guided mode is excited. However, by choosing the appropriate light polarization geometry, the in-plane and out-of-plane tensor components contributing to the TSHG response can be separately monitored. In contrast, an isotropic TR response is negative due to an absorption bleaching process associated with Pauli blocking, thus exposing exclusively the carrier population dynamics mainly through the out-of-plane refractive index modulation as a consequence of a normal incidence pumping geometry applied and the same spot size of the focused pump and probe beams (see the methods section).[48,55,62,63]

Despite the different nature of TR and TSHG, both signals show a rise- and multiple decay-time behavior that can be characterized by the corresponding rise-time ($\tau_R$) and decay-time constants ($\tau_{D1}$, $\tau_{D2}$, and $\tau_{D3}$) and by peak (TR-peak and TSHG-peak), background (TR-BG and TSHG-BG) and unrecovered background (TR-UBG and TSHG-UBG) intensities (Figs. 1 and 2). We assigned the time constants of the TR signals to electron-electron thermalization occurring in both the bulk states and Dirac SS ($\tau_R^{TR}$), electron-longitudinal-optical(LO)-phonon relaxation in the bulk ($\tau_{D1}^{TR}$), which leads to a metastable population of the conduction band edge that continuously feeds a non-equilibrium population of Dirac SS alternatively to the carrier recombination ($\tau_{D2}^{TR}$), and quasi-equilibrium carrier population in Dirac SS ($\tau_{D3}^{TR}$).[48,55,62] The corresponding short time-scale rise- and decay-time constants for the TSHG responses have other physical interpretations, as discussed further below. The TR-peak, TR-BG, TR-UBG, TSHG-peak, TSHG-BG, and TSHG-UBG intensities all increase with increasing probe beam power, but at different rates in accordance with their linear (for TR processes) and quadratic (for TSHG processes) power dependences.[53,62] The TR traces are similar to those reported previously,[48,55,62,63] while the TSHG traces show an oscillatory behavior that is not present in the TR traces.

Specifically, the Fourier transform of the TSHG oscillations observed in the $P_{in}$ - $S_{pump}$ - $P_{out}$ polarization geometry yields a frequency ~42 GHz (~0.174 meV) (Supplementary Fig. 1). Using this frequency, one can fit the oscillatory part to a damped cosine function to yield a ~50 ps damping coefficient (Fig. 1). The oscillations in the TSHG traces measured in the $S_{in}$ - $P_{pump}$ - $S_{out}$ polarization geometry are still present, although they occur at a lower frequency of ~15 GHz (~0.062 meV) (Supplementary Fig. 1) and with a larger damping coefficient of ~80 ps (Fig. 2). In general, both the frequency and damping coefficient of oscillations observed in the $P_{in}$ - $S_{pump}$ - $P_{out}$ polarization geometry can be associated with the coherent acoustic phonon modes. However, these modes of similar frequencies were observed in the TR traces of the films with $d \geq 40$ nm, whereas the frequencies were significantly increased with decreasing $d$ in the range from 40 to 15 nm because the film bulk acoustic wave resonator (FBAWR) modes begin contributing to the dynamics (Fig. 3a). Moreover, the FBAWR modes disappear completely for films with $d < 15$ nm (the range where oscillations in the TSHG traces appear) due to indirect opposite-surface coupling that leads to a gradual elimination of the out-of-plane refractive index modulation as the FBAWR regime crosses over to that of the evanescent guided Lamb wave excitation (Figs. 3a and b).[55]

This change of the interaction regime with decreasing film thickness, which is usually associated with the contribution of the imaginary roots of the Rayleigh-Lamb equation, localize the reactive power which oscillates along the in-plane and out-of-plane directions due to symmetric and antisymmetric displacements of atoms with respect to the mid-plane of the film, respectively.[64] Because of the expected strong coupling of the resulting evanescent fields to Dirac fermions,[36,56] the corresponding TM and TE guided CADP modes can be excited with well-defined acoustic wavevectors and velocities ($\sim 2 \times 10^3$ ms$^{-1}$). It is worth noting that the acoustic velocities are significantly suppressed in this case compared to the Fermi velocity in the TI Bi$_2$Se$_3$ single crystals ($\sim 5 \times 10^5$ ms$^{-1}$) due to the strong indirect opposite-surface coupling regime in the film.[39,55] This kind of coupling between the evanescent fields and Dirac fermions is known as surface plasmon resonance, despite the evanescent fields being created by the evanescent guided Lamb wave instead of the incident light wave as in the more usual Otto/Kretschmann prism couplers or grating structures where the phase-matched plasmon frequency is determined by the incident light wavelength and the grating period, respectively.[44,45] It should also be emphasized here that the low-frequency oscillations observed in the S$_{in}$ - P$_{pump}$ - S$_{out}$ polarization geometry are at least $\sim 2.5$ times lower in frequency than any coherent acoustic phonon modes observed for thin Bi$_2$Se$_3$ films.[55] Thus, oscillations in TSHG traces measured in the P$_{in}$ - S$_{pump}$ - P$_{out}$ and S$_{in}$ - P$_{pump}$ - S$_{out}$ polarization geometries for the 10 nm thick film can be assigned to the modulation of $E_{dc}^{dep}$ by a periodic electric field associated with TE and TM guided CADPs oscillating along the out-of-plane and in-plane directions of the films, respectively. These oscillations should be distinguished from those resulting from the compressive/tensile strains induced by coherent phonons which were assigned to the local breaking of the crystal lattice symmetry[65] or those being due to an electric field modulation resulting from the interference between the incident laser beam and that reflected from propagating acoustic waves.[66] The latter processes are known to appear with identical oscillation frequencies in both TR and TSHG responses.[65,66] In stark contrast, in our case TR oscillations are completely suppressed and only TSHG oscillations were observed for films with $d$ <15 nm, whereas TR oscillations appear for films with $d \geq$15 nm where TSHG oscillations disappear. All these observations unambiguously prove that oscillations in the TSHG traces are not sourced by a modulation of the bulk refractive index as that occurs in TR traces and are solely associated with electron dynamics.

**TSHGRA patterns and depletion electric field screening.** The TSHGRA patterns measured in the $P_{in}$ - $S_{pump}$ - $P_{out}$ polarization geometry reveal a spectacular change of threefold-to-sixfold rotational symmetry with delay time (Fig. 1). The initial threefold rotational symmetry at the TSHG-peak intensity is similar to that observed for the stationary SHG responses from the single-crystalline and thin-film $Bi_2Se_3$ samples measured in the $P_{in}$ - $P_{out}$ polarization geometry.[53,57,58] Because of the multicomponent nature of the SHG response involving hyperpolarizabilities of both the planar hexagon-arranged topmost Se layer[41,49] and Bi-Se bonds arranged into the rhombohedral unit cell along the threefold trigonal axis,[53] the dominating threefold rotational symmetry points out that the depletion-electric-field-induced bulk contribution dominates over the purely surface contribution. This statement is consistent with the corresponding TSHG intensity which can be expressed similarly to the stationary SHG intensity as[53]

$$I_{PSP}(2\omega) = B[c_1 - c_2 \cos(3\varphi) + c_3 \cos^2(3\varphi)], \quad (2)$$

where $B = AI^2(\omega)$ is a proportionality constant for the given experimental conditions, $\varphi$ is the crystal surface rotation angle, and the weight coefficients $c_i$ represent the partial contributions of the isotropic, threefold, and sixfold rotational symmetry components, respectively. It should be noted that a more precise comparison of the TSHGRA patterns for the stationary SHG response and those taken at the TSHG-peak intensity suggests that the threefold rotation symmetry component is even enhanced in the latter case compared to the sixfold one. This enhancement will be discussed further below in more detail. Because the isotropic component is negligible,[53] the observed temporal dynamics of the TSHGRA patterns point to a suppression of the out-of-plane contribution during a characteristic time of ~15 ps when the metastable population of the bulk states becomes high enough to completely screen $E_{dc}^{dep}$, i.e. the exact sixfold rotational symmetry of the TSHG response from the topmost hexagonally arranged Se-Se bonds can be observed.

The rotational symmetry of the TSHGRA patterns measured for the 10 nm thick film in the $S_{in}$ - $P_{pump}$ - $S_{out}$ polarization geometry remains sixfold for any delay times applied and is identical to that observed for stationary SHG measured in the $S_{in}$ - $S_{out}$ polarization geometry (Fig. 2).[53,57,58] The time-independent TSHGRA patterns are consistent with the monocomponent nature of the TSHG response originating exclusively from the in-plane (*xy*) hyperpolarizability of the continuous hexagonal network of Se-Se bonds,[53,57,58]

$$I_{SPS}(2\omega) = Bc_3 \sin^2(3\varphi).  \qquad (3)$$

Because of the same origin of the sixfold rotational symmetry component of TSHGRA patterns measured in both light polarization geometries applied, their relative 30° rotation (Figs. 1 and 2) results from the cofunction identity $\cos^2(3\varphi) = \sin^2[3(\varphi + 30^0)]$ (equations 2 and 3).[53]

We note also here that (i) the single atomic layer nature of the TSHG response eliminates from consideration all the light re-absorption effects on the damping of the guided CADP modes, (ii) similar variations in the SHG rotational symmetry, although with increasing laser power, has been observed for the Si(001)-SiO$_2$ system due to the carrier-induced screening of the dc interfacial electric field,[67] and (iii) the TSHG response from Bi$_2$Se$_3$ single crystals measured in the same polarization geometry was too weak to recognize any temporal changes in TSHGRA patterns.[68] The aforementioned dynamical screening of the depletion electric field and the resulting single atomic layer nature of the TSHG response unambiguously prove that TSHG oscillations are sourced by the Dirac fermion collective excitations.

**A film thickness dependent condition for CADPs excitation.** One of the intriguing findings is a very narrow $d$ range where CADPs can be efficiently excited and the temporal threefold-to-sixfold change of rotational symmetry occurs. Specifically, both effects are maximized for the 10 nm thick film, while being significantly diminished for thicker and thinner films (although still observable for the 8 and 12 nm thick films) (Supplementary Figs. 1, 2 and 3). This behavior agrees with the $d$ dependences of the TSHG-peak, TSHG-BG, and TSHG-UBG intensities (Supplementary Fig. 3) and is consistent with the resonance-type enhancement previously reported for the stationary SHG intensity, which has been associated with the nonlinear excitation of plasmons in indirectly opposite-surface coupled Dirac SS.[53] However, these modes of double layer structures ("optical" and "acoustic"), which are also known for the electrostatically coupled graphene bilayers,[38-40] have much higher frequencies ~2.01 and ~7.56 THz (~8.0 and ~31 meV)[41,49] and should not to be confused with the CADPs driven by strong coupling between Dirac fermions and the evanescent guided Lamb waves discussed here. The coexistence of two kinds of plasmons in the TI $Bi_2Se_3$ films is similar to that occurring on metal surfaces[11,21] and graphene.[16] The resonant enhancement of the stationary SHG and TSHG signals are hence due to high-energy plasmon excitation. In contrast, low-energy CADPs manifest themselves within the carrier relaxation dynamics when some unique thickness dependent condition is reached at which Landau damping of CADPs in $Bi_2Se_3$ films becomes ineffective.

We associate this condition with the fermion tunneling between the opposite-surface Dirac SS, which accompanies the indirect coupling regime for films with $d < 15$ nm and is naturally limited by the direct coupling regime for films with $d < 6$ nm. The fermion tunneling is expected to modify the linear AP dispersion $\omega_{AP}(q) = v_p q$ to

$$\omega_{AP}(q) = \sqrt{\Delta^2 - C_1 q + C_2 q^2}, \qquad (4)$$

where the plasmon energy gap $\Delta = \omega_{AP}(q = 0)$ and the coefficients $C_1$ and $C_2$ nontrivially depend on the electron density, the fermion tunneling amplitude, and the film thickness.[54] It should be especially emphasized here that despite this form of AP dispersion that has been suggested for "acoustic" (antisymmetric) high-energy plasmon mode in the coupled bilayer structures, it can be directly applied to CADPs induced by the evanescent guided Lamb wave because the same dynamical dielectric function characterizes both phenomena. Using the frequencies of the FBAWR modes experimentally observed in these films,[55] and taking into account that the wavevector for the fundamental FBAWR mode is given by $q = 2\pi/\lambda = \pi/d$, where $\lambda = 2d$ is the acoustic wavelength (Fig. 3b), the FBAWR mode dispersion can be constructed as shown in Fig. 3a. The resulting dispersion is linear $\omega_{\text{FBAWR}}(q) = v_p q$ as $q \to 0$, whereas it breaks down for larger $q$ when the evanescent guided Lamb wave regime becomes dominant (Fig. 3b). Assuming that the evanescent guided Lamb waves completely control the wavevectors and frequencies of collective electronic excitations in Dirac SS and taking into account the fermion tunneling between the opposite-surface Dirac SS, the AP dispersion may reveal a minimum.[54] Figure 3a shows a qualitative modeling of the AP dispersion with $\Delta$ = 0.92 meV and coefficients $C_1$ = 5.1 and $C_2$ = 7.684, where the patterned area represents the Landau damping region. Note that the values for $C_1$ and $C_2$ are not necessarily unique and are used here for qualitative purposes only, whereas the value of $\Delta$ is similar to the theoretically predicted value of ~ 1 meV.[54] The energies of the observed CADP modes are therefore expected to occupy the area slightly above the minimum. Although this crude qualitative modelling requires further significant theoretical efforts to obtain a more complete explanation for the existence of CADPs, the model does provide a realistic explanation of why CADPs remain Landau undamped in the very narrow film thickness range centered at $d$ ~10 nm (Supplementary Fig. 1). It is important to note that because the evanescent Lamb wave field is extended over the entire film and therefore affects equally all the free carriers independently wherever they are located (Fig. 3b), all the plasmons excited by the evanescent Lamb wave field are expected to be Landau damped on the very short time-scale of a few tens of fs, except for Dirac plasmons that remain undamped due to fermion tunneling between the opposite-surface Dirac SS (Fig. 3a). When fermion tunneling becomes inefficient for films thinner that ~8 nm and thicker than ~12 nm, Dirac plasmons are also damped. Because the TSHG technique that we used is sensitive exclusively to Dirac SS, and because the fermion tunneling condition uniquely occurs in a certain thickness range of $Bi_2Se_3$ films, the observation of CADPs becomes possible. Although the resonance-type acoustic-plasmon-to-acoustic-phonon coupling has been considered theoretically only for Bi so far,[36] we assume here that similar strong coupling may occur in $Bi_2Se_3$ as well. At the very least, our experimental results clearly point out that this strong coupling does exist in a similar way as that observed in complex nanostructures.[56] We note also that because of the unusual behavior of the 10 nm thick film, we grew a second sample using the same growth conditions, which also had similar resonant characteristics.

**Ultrafast carrier SS-bulk-SS vertical transport.** The simultaneous measurements of the TR and TSHG signals allow for a comprehensive study of the ultrafast carrier dynamics in thin films of the TI $Bi_2Se_3$, which shed light on the aforementioned screening of $E_{dc}^{dep}$ and conditions under which CADPs can be excited. Figure 3c shows the normalized TR and TSHG traces for the 10 nm thick film. As we discussed above, the TR signal increases with $\tau_R^{TR}$ ~0.3 ps due to the electron-electron thermalization and subsequently decreases with $\tau_{D1}^{TR}$ ~2.15 ps due to the electron-LO-phonon scattering.[48,55,62,63] In contrast, the TSHG signal increases more slowly with $\tau_{R1}^{SHG}$ ~1.1 ps and decays faster with $\tau_{D1}^{SHG}$ ~1.6 ps. Because the optical excitation used (~$10^{20}$ cm$^{-3}$)[62] exceeds the natural *n*-doping level in the films (~$10^{19}$ cm$^{-3}$),[41] the initial $E_{dc}^{dep}$, which originates from the upward band bending at the surface due to the higher density of free electrons residing in Dirac SS than those in the bulk states,[59] can be screened/enhanced depending on a balance of photoexcited electrons between the bulk states and Dirac SS. Specifically, due to the direct optical coupling of incident 1.51 eV photons to Dirac SS,[69] electrons below the Fermi level of occupied Dirac SS (1SS) can be photoexcited into the unoccupied Dirac SS (2SS) with an efficiency comparable to or even higher than that of the common valence-to-conduction band transitions in the bulk (Fig. 3d). The longer $\tau_{R1}^{SHG}$ compared to $\tau_R^{TR}$ confirms that the screening/enhancement dynamics of $E_{dc}^{dep}$ develop only after the electron-electron thermalization establishes a Fermi-Dirac distribution of photoexcited electrons. Due to an overlap in energy between Dirac 2SS and the high energy bulk bands, the thermalized electrons initially populate the upper cone of Dirac 2SS with high efficiency,[48] thus enhancing the initial $E_{dc}^{dep}$, which can be presented by solving the corresponding Poisson equation,

$$E_{dc}^{dep} = -\frac{ez_{dep}^2}{\varepsilon\varepsilon_0}\left(\frac{n_{SS}}{z_{SS}} - \frac{n_B}{z_{dep}}\right), \tag{5}$$

where $z_{SS}$ and $z_{dep}$ are the widths of the Dirac SS actual range (~3 nm – a half of the critical film thickness for direct opposite-surface coupling)[51] and the surface depletion layer (~6.5 nm),[48,59] respectively; $n_{SS}$ and $n_B$ are the densities of electrons residing in Dirac SS and the bulk states, respectively, and $\varepsilon$ is the effective low-frequency dielectric constant. The enhancement of $E_{dc}^{dep}$ appears as an initial rise of the TSHG signal. The subsequent decay of the TSHG response can hence be associated with $E_{dc}^{dep}$ screening due to the spatial redistribution of photoexcited electrons towards the bulk, which occurs primarily through electron-LO-phonon scattering. The resulting increase of the bulk electron density finally changes $E_{dc}^{dep}$ sign (Eq. 5), the process which is accompanied by the change of the initial upward band bending to the downward band bending that permits 2D electron gas (2DEG) to coexist with an electron population in Dirac 1SS.[61] These dynamics appear as a drop of the TSHG signal to the minimal intensity at the dip which is controlled by the total density of photoexcited electrons dynamically residing in the bulk. This behavior is consistent with the temporal threefold-to-sixfold change of rotational symmetry in the TSHGRA patterns (weight coefficient $c_2$ in Eq. 2) that closely follows the decay trend of the TSHG response (Fig. 3c). The minimal TSHG intensity, which can be characterized by the ratio of the SHG intensity at ~7 ps (a dip) to that at ~1.7 ps (TSHG-peak), progressively increases with decreasing $d$ from 12 to 6 nm for the $P_{in}$ - $S_{pump}$ - $P_{out}$ polarization geometry, whereas it remains almost constant for the $S_{in}$ - $P_{pump}$ - $S_{out}$ polarization geometry (Supplementary Fig. 3). This behavior agrees with an increase of the electron-hole recombination rate in Dirac SS with decreasing $d$, because the density of electrons dynamically residing in the bulk is inversely proportional to the rate of carrier recombination in Dirac SS.[48]

After the TSHG intensity dips, it increases again with a second rise-time of $\tau_{R2}^{SHG}$ ~5.8 ps. Taking into account the scattering processes between 2DEG and Dirac 1SS,[61] the second rise of the TSHG signal and the corresponding second peak indicate the balancing of electron density between the 2DEG and Dirac 1SS by electron transfer from the former to the latter (Fig. 3d).[48,55,62] This dynamical increase of electron density in Dirac 1SS progressively weakens $E_{dc}^{dep}$ to 0, the process which is self-consistently controlled by the 2DEG-to-Dirac-1SS electron scattering rate. Correspondingly, the weight coefficient $c_2$ is stabilized at the minimal level, indicating that the purely surface second-order susceptibility tensor dominates the TSHG response. Subsequently, the TR and TSHG signals decrease with longer decay-times of $\tau_{D2}^{SHG} \approx \tau_{D2}^{TR}$ ~250-280 ps. Because both signals reveal similar relaxation dynamics at this stage, we associate this decay with carrier recombination in Dirac SS.[48]

The discussed ultrafast carrier SS-bulk-SS vertical transport and the resulting quasi-equilibrium between electron populations in 2DEG and Dirac 1SS also establishes the unique and stringent condition for the excitation of CADPs by the evanescent guided Lamb waves developed within the same relaxation process and for the fermion tunneling between the opposite-surface Dirac SS through the region where the massless fermions acquire a finite mass and hence prevents CADPs from Landau damping similarly to the coupled bilayer structures.[54] A subsequent much longer decay ($\tau_{D3}^{SHG}$) of the TSHG signal ranges a few ns and is similar to $\tau_{D3}^{TR}$. Consequently, both these long decay-time features and the corresponding TR-UBG and TSHG-UBG intensities characterize the quasi-equilibrium electron population in Dirac 1SS. Finally, TR-BG and TSHG-BG intensities are likely due to a capacitor-type electric field ($E_{dc}^{cap}$) developed in the film because $\tau_{D3}^{TR}$ and $\tau_{D3}^{SHG}$ are almost comparable to the inverse of the repetition rate of the laser used (12.5 ns), thus allowing a quasi-steady accumulation of photoexcited electrons in Dirac 1SS as long as the sample is illuminated.[48,53,55,62]

**Discussion**

The observations presented in this article allow us to highlight several approaches which would be interesting to a broad audience of scholars exploring coherent low-energy collective electronic excitations in topological insulators. One of them is the application of nonlinear optical techniques allowing avoiding the photon-plasmon momentum mismatch for the excitation of surface plasmons. This approach seems to provide an advantage compared to the linear optical techniques, for which the waveguide evanescent Otto/Kretschmann prism couplers or grating structures deposited on the surface of topological insulators are required. In the latter case, which is widely used, the phase-matched plasmon frequency is limited only to the range determined by the micro-ribbon width and the grating period.[46,47] Moreover, the deposition of the micro-ribbon grating can hypothetically break the time-reversal symmetry of massless fermions in Dirac SS and hence strongly affect not only the intrinsic plasmon dynamics in the films, but also a unique metallic-type nature of Dirac SS, turning them into a trivial insulator state with massive fermions.

As we mentioned here, there are only a few reports discussing a time-domain detection of the slab thickness dependent oscillations which were associated with coherent acoustic plasmons.[32,33] However, these oscillations were detected using photocurrent autocorrelation and TR techniques, both being incapable, in principle, of distinguishing between the surface and bulk contributions and hence allowing for various interpretations. This statement is extremely important for topological insulators and Dirac systems in general, where the surface contribution is of major interest. Therefore, only applying surface and depletion electric field sensitive techniques, such as TSHG, the unambiguous time-domain detection of CADPs in topological insulators seems to be possible. Moreover, we have demonstrated here an advantage of simultaneous measurements of TR and TSHG responses, which due to their different origins provide us with more comprehensive information about carrier dynamics in topological insulators.

In this article we have also demonstrated a successful application of TSHGRA technique which allows for monitoring the crystal rotational symmetry with delay time. The observed spectacular change of threefold-to-sixfold rotational symmetry with delay time resulting from the dynamical screening of the surface depletion electric field in the TI Bi$_2$Se$_3$ is unique and has never been reported previously. One of the attempts to use this technique was less successful due to extremely noisy SHG signals.[68]

Using TSHGRA technique, we have demonstrated that the dynamical elimination of the depletion-field-induced SHG response allows for observation of the exact sixfold rotational symmetry associated with the topmost hexagonally arranged Se-Se bonds in Bi$_2$Se$_3$ films. The corresponding oscillations observed in this regime hence unambiguously characterize CADPs. The sensitivity of this technique exceeds that of the time-resolved and angle-resolved photoemission spectroscopy (TrARPES), which is known to monitor a few nanometers range in close proximity to the surface.

Another remarkable finding discussed in our article is the experimental proof of the validity of the theoretical prediction made by Das Sarma and Hwang[54] regarding plasmon excitation in coupled bilayer structures, which includes the modification of the acoustic plasmon dispersion when fermion tunneling between the opposite-surface Dirac SS arises with decreasing film thickness. The effect resonantly appears for the film thickness range limited by that associated with the direct coupling between the opposite-surface Dirac SS (6 nm) and that associated with the indirect coupling (15 nm). This resonance-type behavior in Bi$_2$Se$_3$ films centered at ~10 nm is consistent with those appeared in the enhancement of high-frequency THz-range optical phonon modes,[41] nonlinear susceptibilities,[53] electron-phonon scattering strength,[49] and electron-phonon relaxation dynamics.[48,49,62] All these observations point out that indirect opposite-surface coupling has a significant impact on carrier dynamics in this thickness range of the thin-film topological insulator Bi$_2$Se$_3$.

Finally, we note that the CADP modes observed in Bi$_2$Se$_3$ thin films using the TSHG technique could be expected across a wide range of 2D materials because the excitation of acoustic phonons and electron tunneling are common stages of various electronic processes occurring in these materials.

## Methods

**Samples.** Bi$_2$Se$_3$ thin-films of 6, 8, 10, 12, 15, 20, 25, 30, 35, and 40 nm thick were grown on 0.5 mm Al$_2$O$_3$(0001) substrates by molecular beam epitaxy (MBE), with a 10 nm thick MgF$_2$ protecting capping layer, which was grown at room temperature without exposing the film to atmosphere. The quality of the films, their structure and thicknesses were determined from x-ray reflectivity, reflection high energy electron diffraction (RHEED), and x-ray diffraction (XRD) measurements.[49] The Hall conductivity measurements performed at 2 K showed that all films have *n*-doping level in the range of $0.5 - 3.5 \times 10^{19}$ cm$^{-3}$.[41]

**Experiments.** The TR and TSHG measurements were performed using a multifunctional pump-probe setup that includes a Ti:Sapphire laser with an output average power of the fundamental laser beam of 2.5 W, pulse duration $\tau_L$ = 100 fs, center photon energy 1.51 eV and repetition rate 80 MHz.[53] The pump beam was at normal incidence and the probe beam was at an incident angle of ~15°, focused through the same lens to a spot diameter of ~100 μm. A pump beam average power was 640 mW. A probe beam average power was in the range of 100 – 580 mW. The TR and TSHG responses were measured simultaneously in the reflection geometry with various step sizes in delay times between the pump and probe pulses using a photodiode and photomultiplier tube, respectively. The signals were collected with a lock-in amplifier triggered at the pump beam modulation frequency of 800 Hz. The TSHGRA measurements were performed at various pump-to-probe delay times by rotating the sample mounted on a rotation stage about the surface normal with a step size of 2°. The linear polarization of the beams was either P (in the plane of incidence) or S (in the plane of the film). Four different light polarization geometries for the incident laser beam (in), the outgoing reflected fundamental and SHG beams (out), and the normal incidence pump beam were used to measure TSHGRA patterns: P$_{in}$ - S$_{pump}$ - P$_{out}$, S$_{in}$ - S$_{pump}$ - P$_{out}$, P$_{in}$ - P$_{pump}$ - S$_{out}$, and S$_{in}$ - P$_{pump}$ - S$_{out}$. Experiments were performed in air and at room temperature. No film damage was observed for the laser powers used in the measurements reported here.

**Acknowledgements**

This work was supported by a Research Challenge Grant from the West Virginia Higher Education Policy Commission (HEPC.dsr.12.29). Some of the work was performed using the West Virginia University Shared Research Facilities.

**Author contributions**

Y.D.G. created the optical setup and conducted the optical experiments. $Bi_2Se_3$ films were MBE grown and characterized using high resolution x-ray reflectivity by S.B. T.A.J coated the $Bi_2Se_3$ films with the $MgF_2$ layer. The measurements were performed in the laboratory hosted by M.B.H. D.L. guided the research, supervised the sample growth and characterization, and assisted with the manuscript preparation. All authors contributed to discussions. Y.D.G. prepared the manuscript.

Figure captions.

**Figure 1 I TR and TSHG traces for the 10 nm thick Bi$_2$Se$_3$ film.** TR and TSHG traces for the 10 nm Bi$_2$Se$_3$ film simultaneously measured with 250 fs step-size in the P$_{in}$ - S$_{pump}$ - P$_{out}$ polarization geometry at different pump/probe powers as indicated by the corresponding colors. The assignment of TR-peak, TSHG-peak, TR-BG, SHG-BG, TR-UBG and SHG-UBG intensities are shown. The normalized TR rotational anisotropy and TSHGRA patterns measured at the fixed pump-to-probe delay times as indicated by arrows are shown in insets of the corresponding TR and TSHG sections of the graph. The red curves present the best fit to the data. The TSHG section also indicates the frequency and damping coefficient of the oscillatory part of the TSHG trace.

**Figure 2 I TR and TSHG traces for the 10 nm thick Bi$_2$Se$_3$ film.** Upper panel shows TR and TSHG traces for the 10 nm Bi$_2$Se$_3$ film simultaneously measured with 250 fs step-size in the S$_{in}$ - P$_{pump}$ - S$_{out}$ polarization geometry at different pump/probe powers as indicated by the corresponding colors. The TSHG section also indicates the frequency and damping coefficient of the oscillatory part of the TSHG trace. Lower panel shows the corresponding longer delay-time range TSHG trace measured with 1 ps step-size at 640/580 mW pump/probe powers. The normalized TSHGRA patterns measured at the fixed pump-to-probe delay times as indicated by arrows are shown in the corresponding insets. The red curves in both panels present the best fit to the data.

**Figure 3 I CADP modes and the unique conditions for their excitation in thin-films of Bi$_2$Se$_3$.** (a) Dispersion of the FBAWR modes observed in Ref. 55 for the same Bi$_2$Se$_3$ films of different thicknesses (red dots) and the excited TE and TM guided CADP modes (green and blue dots, respectively) are shown. Red dashed line approximates the linear dispersion of FBAWR modes,

whereas solid black line shows the dispersion of AP in the presence of fermion tunneling between the opposite-surface Dirac SS, which was modeled using equation 4 with $\Delta$ = 0.92 meV, $C_1$ = 5.1, and $C_2$ = 7.684. Patterned area represents the Landau damping regions. (b) A schematic representation of the FBAWR modes for films with $d >$ 15 nm and the extensional (symmetric) and flexural (antisymmetric) Lamb modes for films with $d <$ 15 nm. Blue arrows indicate the direction of the evanescent guided Lamb wave oscillations. (c) The normalized TSHG and negative TR traces for the 10 nm thick film are shown together with the fit curves (black solid lines). To guide the general tendency, the fit to the TSHG trace is shown without the oscillatory part. Inset shows zoom in on the same traces. The rise- and decay-time constants are shown in the corresponding colors. Black squares represent the temporal dynamics of the threefold symmetry component of the TSHGRA patterns (weight coefficient $c_2$ in equation 2). (d) Schematic sketch of the electronic structure of $Bi_2Se_3$ films and the relaxation dynamics in Dirac SS due to their direct optical coupling to ~1.51 eV incident photons (thicker red arrows). Thinner-red and multicolor arrows present the recombination and LO-phonon-assisted relaxation processes, respectively. Black dashed arrows indicate 2DEG-to-Dirac1SS scattering which balances the corresponding electron densities and completely eliminates the depletion electric field ( $E_{dc}^{dep}$ = 0), thus leading to the dominant contribution of purely surface sixfold rotational symmetry component originating from the topmost Se atomic layer of the film.

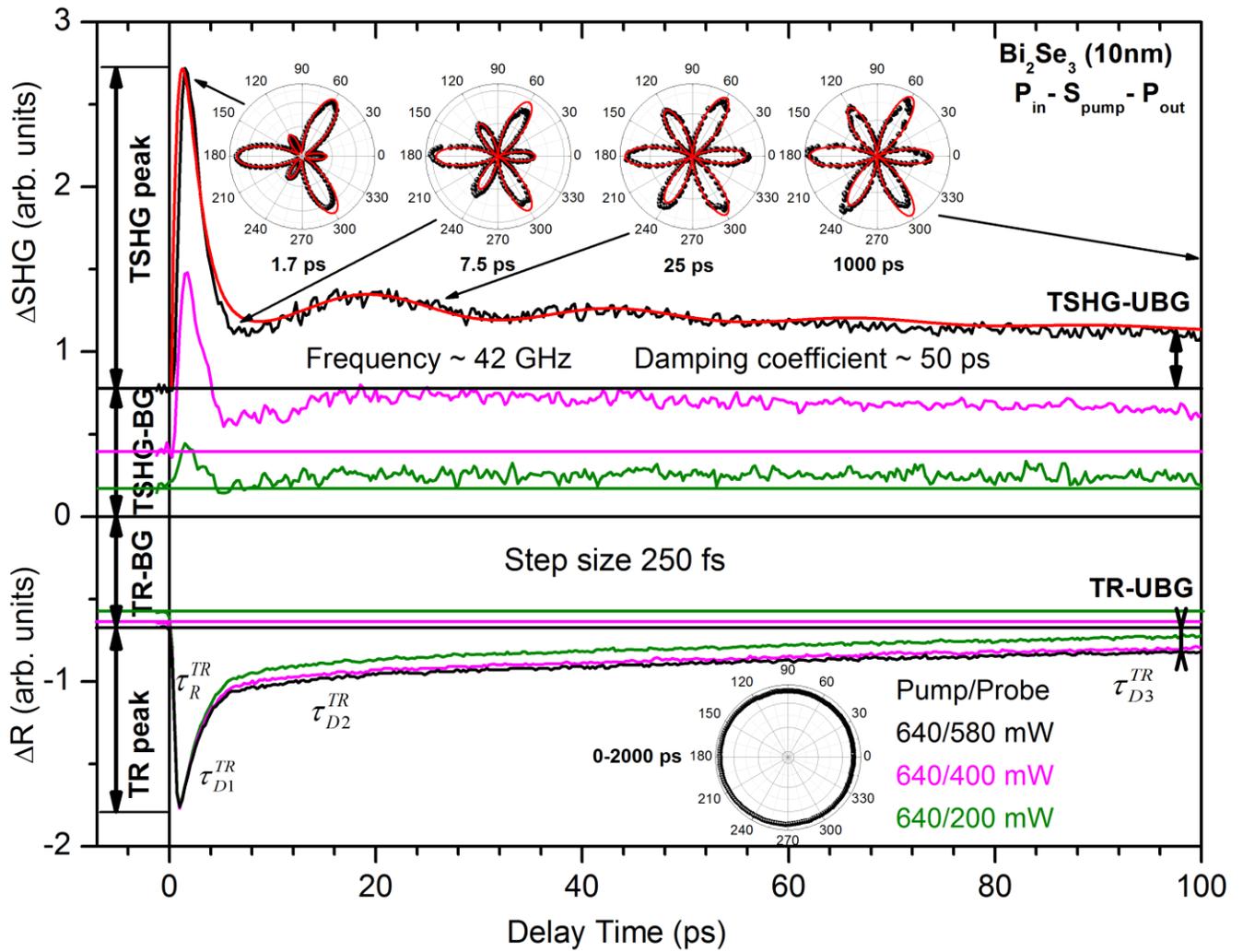

**Figure 1**

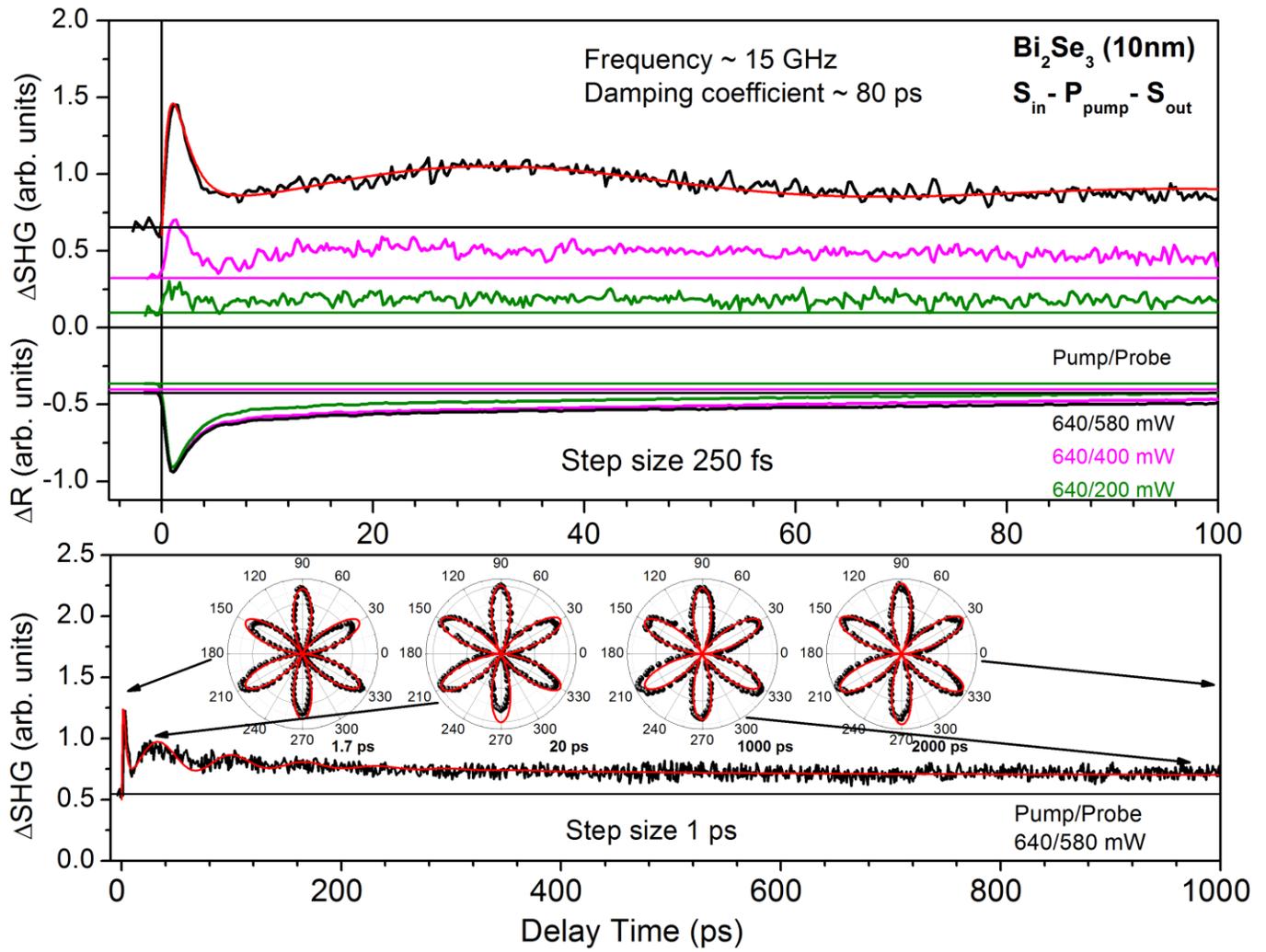

**Figure 2**

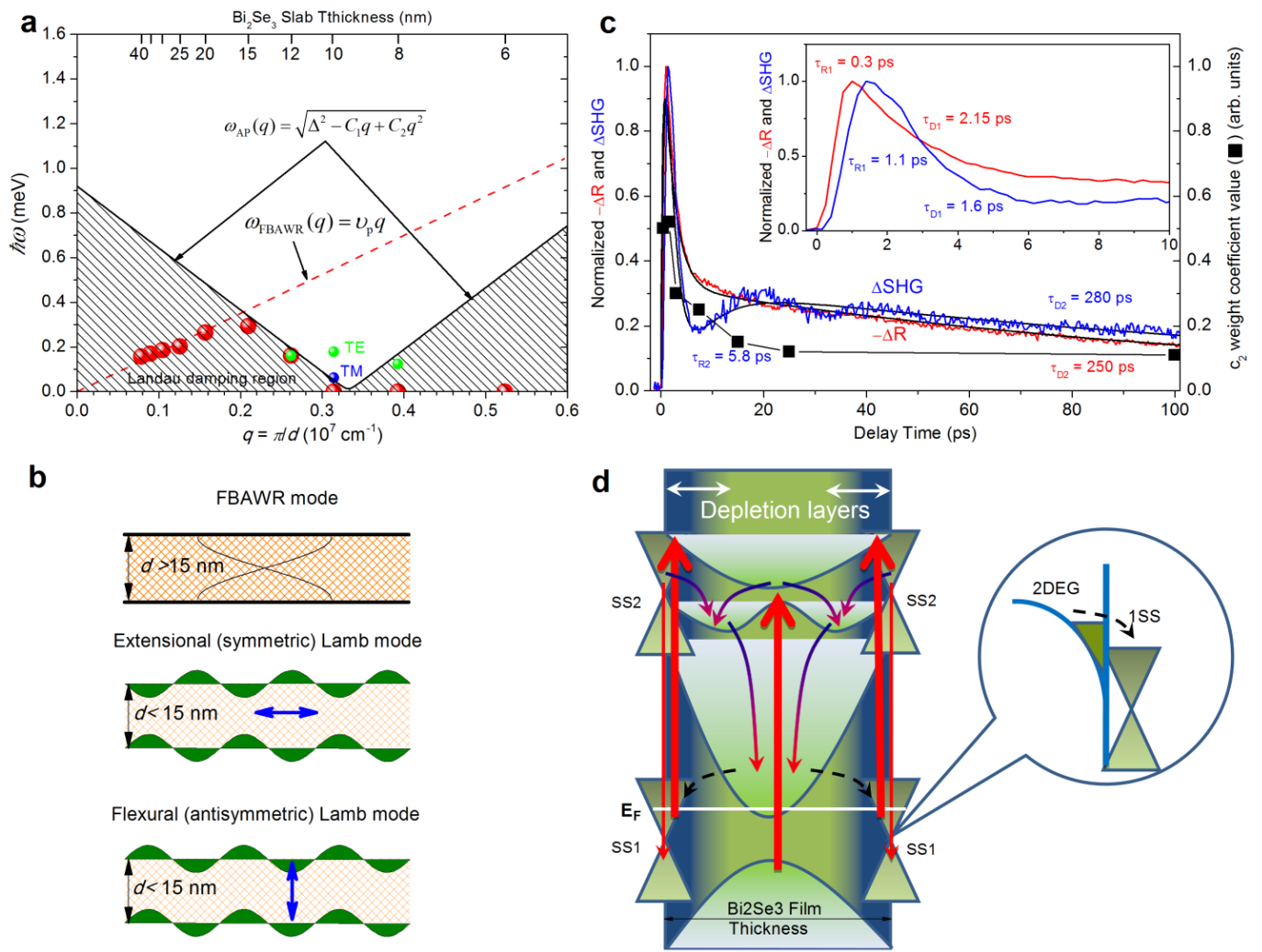

**Figure 3**

# Supplementary Information

**Supplementary Figures**

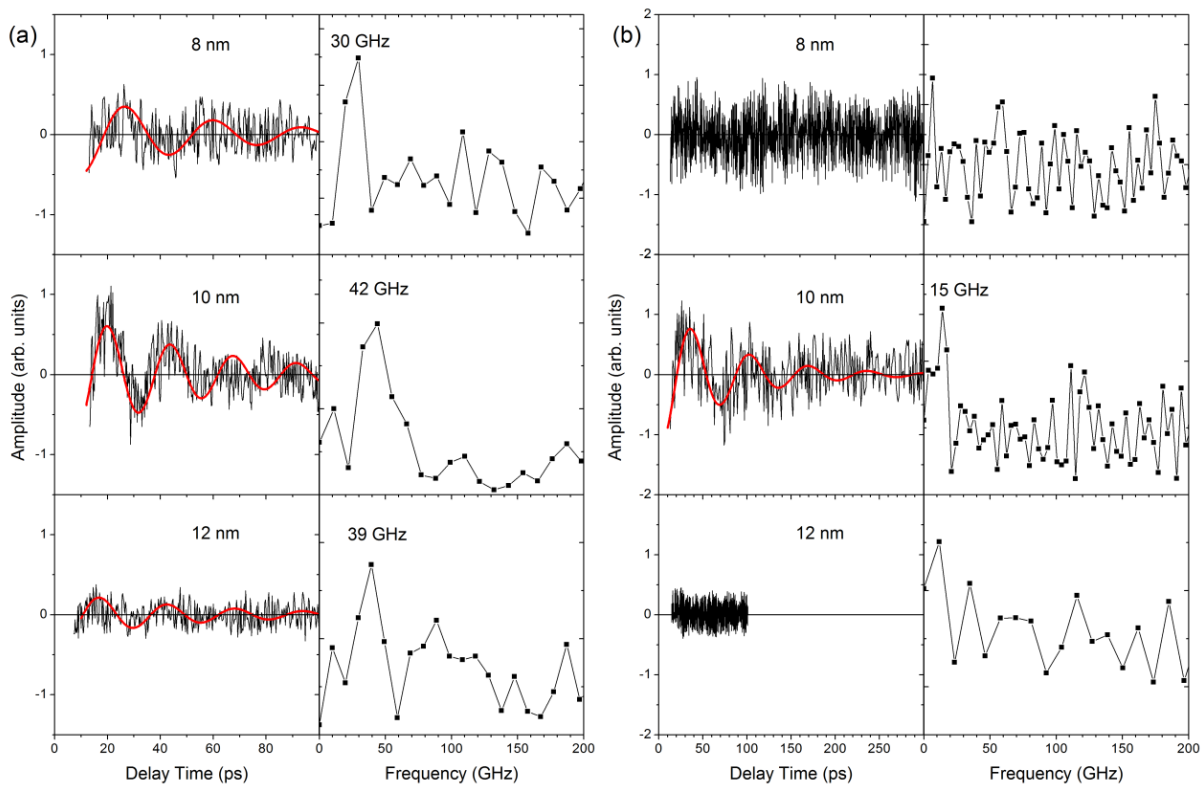

**Supplementary Figure 1 | An analyses of the oscillatory part of the TSHG traces.** The extracted oscillatory part of the TSHG traces (left columns) and their fast Fourier transforms (FFT) (right columns) for $Bi_2Se_3$ films of various thicknesses as indicated in nm, which were measured in the $P_{in}$ - $S_{pump}$ - $P_{out}$ (a) and $S_{in}$ - $P_{pump}$ - $S_{out}$ (b) polarization geometries. Figure (a) clearly demonstrates ~42 GHz oscillations measured in the $P_{in}$ - $S_{pump}$ - $P_{out}$ polarization geometry for the 10 nm thick film, which however become significantly suppressed (although still observable) and show slightly decreased frequencies for thinner and thicker films. Alternatively, ~15 GHz oscillations which were measured in the $S_{in}$ - $P_{pump}$ - $S_{out}$ light polarization geometry for the 10 nm thick film are completely damped for thinner and thicker films (the corresponding FFT curves show just a noise level) [Figure (b)]. Red curves present the result of the fit to the damped cosine function using FFT frequencies and damping coefficients 50 ps (a) and 80 ps (b).

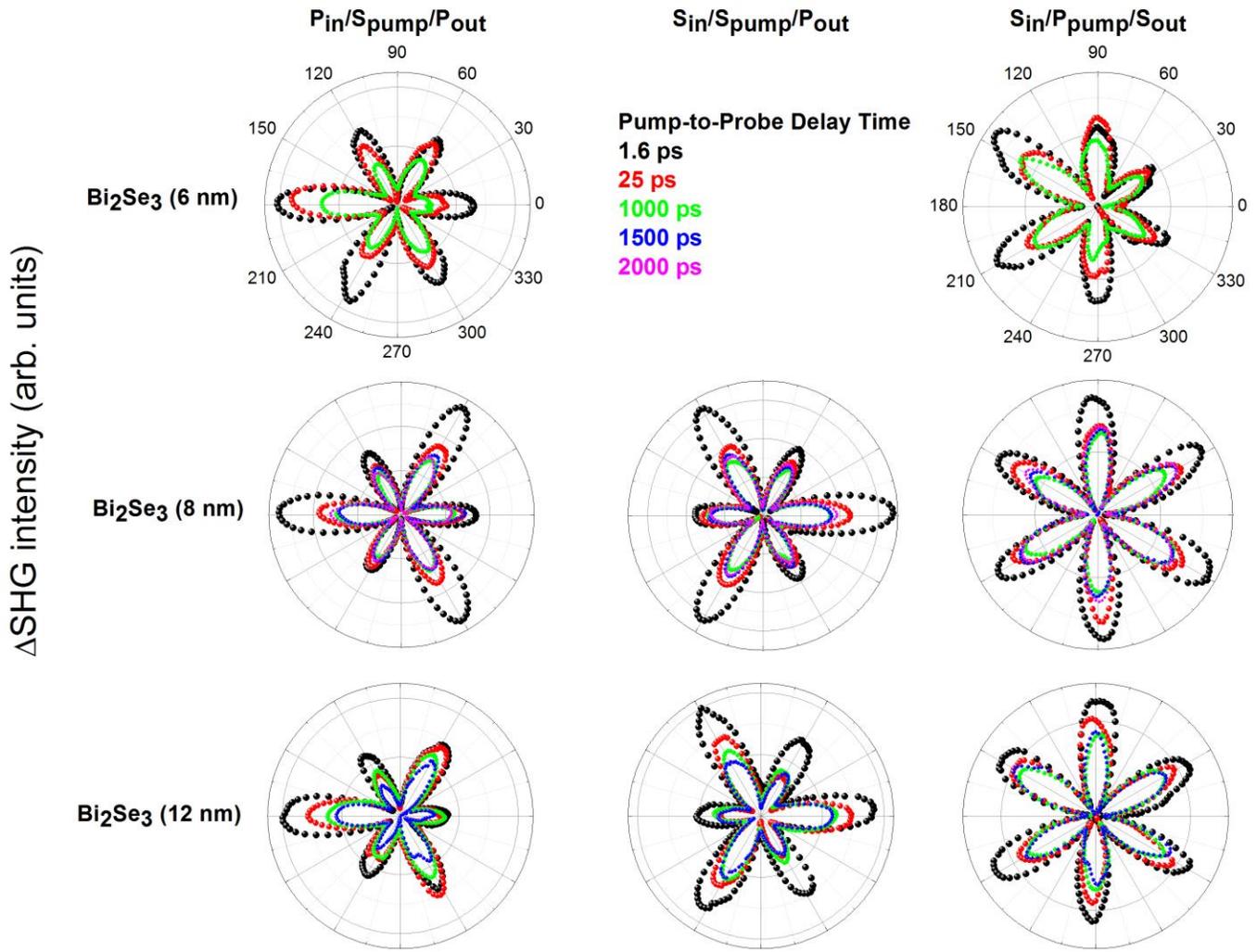

**Supplementary Figure 2 | TSHGRA patterns for Bi$_2$Se$_3$ films of various thicknesses.** The TSHGRA patterns of the 6, 8, and 12 nm thick Bi$_2$Se$_3$ films (rows) measured in the P$_{in}$ - S$_{pump}$ - P$_{out}$, S$_{in}$ - S$_{pump}$ - P$_{out}$, and S$_{in}$ - P$_{pump}$ - S$_{out}$ light polarization geometries (columns) at different pump-to-probe delay times as indicated by the corresponding colors. The TSHGRA patterns of the 6 nm thick film reveal a significant distortion of the rotational symmetry in a similar way as that observed for the stationary SHG response [Ref. 52 in the paper]. The TSHGRA patterns measured in the S$_{in}$ - P$_{pump}$ - S$_{out}$ light polarization geometry do not show any temporal changes in rotational symmetry for the films of various thicknesses, similar to the experimental findings observed for the 10 nm thick film (see figure 2 in the original paper). In contrast, the threefold-to-sixfold symmetry change of the TSHGRA patterns can be observed in the P$_{in}$ - S$_{pump}$ - P$_{out}$ and S$_{in}$ - S$_{pump}$ - P$_{out}$ light polarization geometries. The largest effect is for the 10 nm thick film (see figure 1 in the original paper), smaller for the 8 nm thick film, and is significantly suppressed for the films with $d <8$ nm and $d >10$ nm.

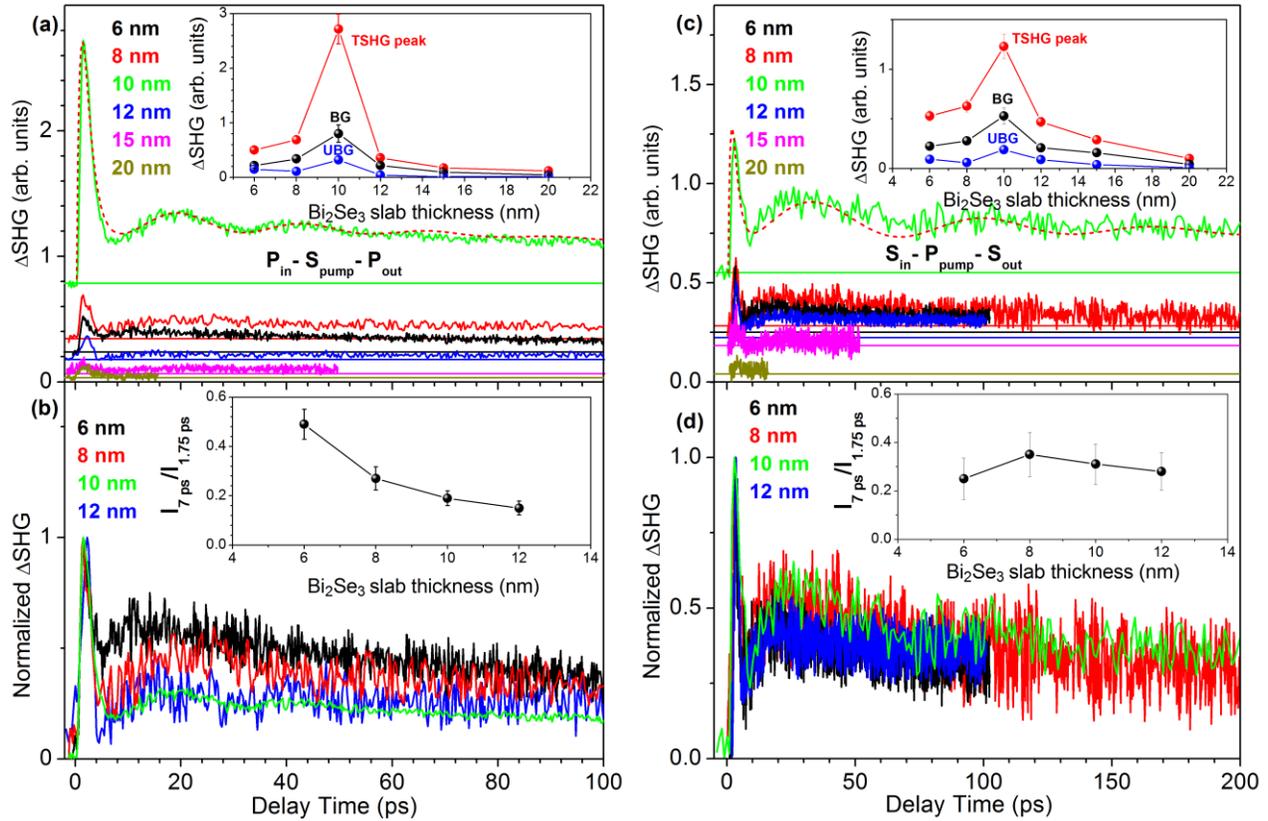

**Supplementary Figure 3 | TSHG traces for $Bi_2Se_3$ films of various thicknesses.** The TSHG traces of $Bi_2Se_3$ films of various thicknesses as indicated by the corresponding colors measured in the $P_{in}$ - $S_{pump}$ - $P_{out}$ [(a) and (b)] and $S_{in}$ - $P_{pump}$ - $S_{out}$ [(c) and (d)] polarization geometries. Figures (b) and (d) show the same curves shown in (a) and (c), but being normalized. The TSHG traces shown in (a) and (c) for the 10 nm thick films and red dotted curves presenting the result of the fit are the same as those shown in figures 1 and 2 of the original paper. Insets in (a) and (c) present the $d$ dependences of the TSHG-peak, TSHG-BG, and TSHG-UBG intensities. Insets in (b) and (d) show the $d$ dependences of the ratio of the intensities at ~7 ps (a dip of the TSHG traces) to that at ~1.7 ps (TSHG-peak). One can clearly see that the oscillatory part is superimposed on the intensity of the second peak of the TSHG response. Consequently, the oscillatory part intensity is significantly suppressed for the films with $d$ <8 nm and $d$ >10 nm, despite the second peak of the TSHG response is always present.